\documentclass[journal]{IEEEtran}
\usepackage{amsmath,amssymb}
\usepackage{graphicx}
\usepackage{array}
\usepackage{url}
\usepackage{balance}
\usepackage[noadjust]{cite}
\usepackage{xcolor}
\usepackage{enumerate}
\usepackage{makecell}
\newtheorem{thm}{Theorem}

\newtheorem{rem}{Remark}
\begin{document}

\title{Secretive Coded Caching with Shared Caches}

\author{Shreya Shrestha Meel,
        and~B.~Sundar~Rajan,~\IEEEmembership{Fellow,~IEEE}
\thanks{Manuscript received ; revised . This work was supported partly by the Science and Engineering Research Board (SERB) of Department of Science and Technology (DST), Government of India, through J. C. Bose National Fellowship to B. Sundar Rajan.}
\thanks{The authors are with the Department of Electrical Communication Engineering, Indian Institute of Science, Bangalore 560012, India (e-mail:
\{shreyameel, bsrajan\}@iisc.ac.in).}
}


\maketitle

\begin{abstract}
We consider the problem of \emph{secretive coded caching} in a shared cache setup where the number of users accessing a particular \emph{helper cache} is more than one, and every user can access exactly one helper cache. In secretive coded caching, the constraint of \emph{perfect secrecy} must be satisfied. It requires that the users should not gain, either from their caches or from the transmissions, any information about the content of the files that they did not request from the server. In order to accommodate the secrecy constraint, our problem setup requires, in addition to a helper cache, a dedicated \emph{user cache} of minimum capacity of 1 unit to every user. This is where our formulation differs from the original work on shared caches (``Fundamental Limits of Coded Caching With Multiple Antennas, Shared Caches and Uncoded Prefetching'' by E.~Parrinello,  A.~{\"{U}}nsal and P.~Elia in Trans. Inf. Theory, 2020). In this work, we propose a secretively achievable coded caching scheme with shared caches under centralized placement. When our scheme is applied to the dedicated cache setting, it matches the scheme by Ravindrakumar \emph{et al.} (``Private Coded Caching'', in Trans. Inf. Forensics and Security, 2018).
\end{abstract}

\IEEEpeerreviewmaketitle

\section{Introduction}\label{sec1}
\IEEEPARstart{I}{n} wireless content delivery networks, caching helps in reducing the disparity in traffic load between peak and off-peak hours, by allowing end users to store parts of the library in their local caches. The seminal work by Maddah-Ali and Niesen in \cite{maddah2014fundamental} established that congestion of the shared broadcast link can be reduced further by making coded transmissions that are useful to more than one user simultaneously. This system operates in two phases-\emph{placement phase} and \emph{delivery phase}. In the placement phase, the server fills the users' caches with some fraction of \emph{all the files}, without knowing their future demands. In the delivery phase, users reveal their file requests, and the server responds by sending coded multicast transmissions over the broadcast link such that, the users, from the transmissions and cache contents, can recover their requested files. The number of bits transmitted over the shared link, normalized by the file size, is known as the \emph{rate} of the coded caching scheme. To reduce the load on the link, rate needs to be minimised. However, the placement and delivery as in \cite{maddah2014fundamental} violates the secrecy constraint, as users get some information about all the files from their caches, even before they reveal their demands. In secretive coded caching, the two phases must be designed such that, users obtain \emph{no information} about the files that they do not request. 

Secretive coded caching problem was first studied in \cite{ravindrakumar2016fundamental,ravindrakumar2018private} where the authors proposed an achievable coded caching scheme imposing the constraint of perfect secrecy, under both centralized and decentralized settings. Their general scheme was also secure against eavesdroppers, as in \cite{rasengupta2015fundamental}. This problem was extended to the setup of Device-to-Device caching networks in \cite{zewail2020device} and to Combination networks in \cite{zewail2018combination}. The authors in \cite{ma2019secure} considered secretive coded caching in a setting where at most $l$ out of the $K$ users are untrustworthy, and might collude among themselves. On setting the value of $l$ to unity, this scheme reduces to the original scheme in \cite{ravindrakumar2018private}. 

All the works cited above \cite{ravindrakumar2016fundamental,ravindrakumar2018private,zewail2020device,zewail2018combination, ma2019secure}, consider a dedicated caching network, i.e., each user is privileged to access a cache, distinct from the remaining users in the network. However, to the best of our knowledge, in a shared cache setup, where multiple users are assisted by a single helper cache, secretive coded caching has not been studied so far. References \cite{parrinello2020fundamental} and \cite{ibrahim2019coded} studied coded caching in this setting, without imposing the secrecy constraint, with uncoded and coded placement respectively. In \cite{parrinello2020fundamental}, the authors formulate the worst case achievable rate (which they refer to as `delay') and prove its optimality under uncoded placement through an index-coding-aided converse. They create multicasting opportunities in their delivery algorithm, by leveraging the content overlap between distinct caches, similar to that in \cite{maddah2014fundamental}. In \cite{ibrahim2019coded}, the authors provide a scheme achieving lower worst-case delay than \cite{parrinello2020fundamental} by placing uncoded and coded sub-files in the helper caches, utilizing the knowledge of the user to cache association profile and optimising over the size of these sub-files to minimise rate.

Like any secretive coded caching problem, in our shared cache setup too, secrecy of content of files has to be preserved in two aspects: (i) caching, and (ii) transmission. To ensure no leakage of information via (i), we employ an appropriate \emph{secret-sharing scheme} to encode the files before they are placed at the \emph{helper caches}. By encoding the files this way, the cached contents alone reveal no information about the files to the users. For no information leakage via (ii), we securely deliver each multicast message so that, only the users who benefit from the transmission can decode it. This is done by XOR'ing each message with a unique \emph{key}. This necessitates that the key used for encoding  each multicast message serving a group of users is made privately available to the \emph{user caches} of this group. Otherwise, any user associated to a given cache, can recover the files requested by the other users connected to the same cache, resulting in a breach of the secrecy constraint. We outline the contributions of this letter below:
\begin{itemize}
    \item We provide a secretively achievable coded caching scheme for a caching system with shared caches. 
    \item When our scheme is applied to the dedicated caches setup, we show that the secretively achievable rate-memory pair as obtained in \cite{ravindrakumar2018private} is recovered. 
\end{itemize}

The rest of the paper is organized as follows. Section \ref{sec2} sets up the problem. The proposed scheme is presented in Section \ref{sec3}. In Section \ref{sec4}, we discuss the main results of the paper.  Finally, we conclude the paper with Section \ref{sec5}.
\section{Problem Setup}
\label{sec2}
\begin{figure}[htbp]
    \centering
   \vspace*{-3.5cm}
   \includegraphics[width=4in]{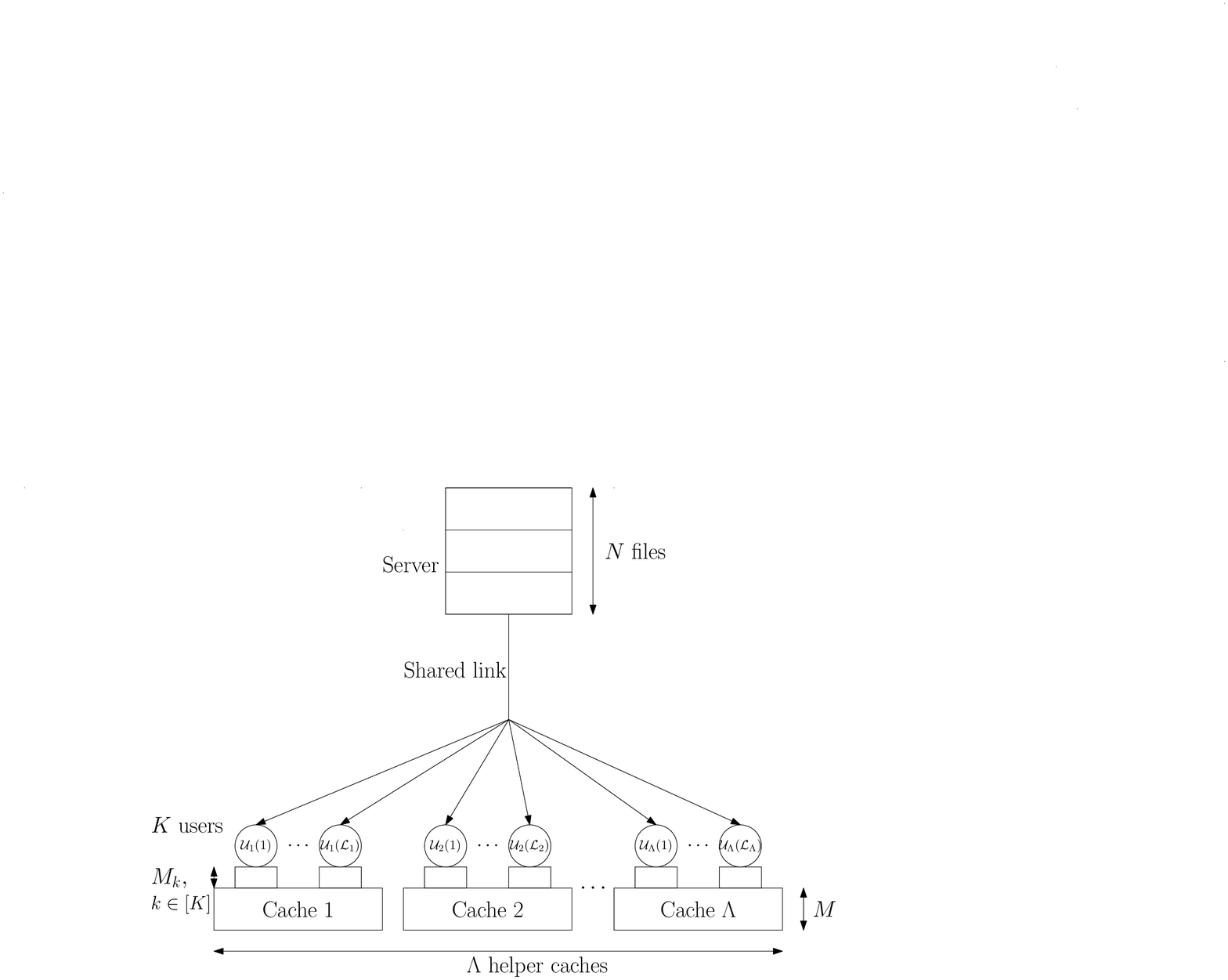}
   
    \caption{Caching system model imposing secrecy constraint.}
    \label{fig_1}
\end{figure}

The system model is illustrated in Fig. \ref{fig_1}. The server has a library of $N$ files, denoted by $W^{[N]}:=\{W^1,W^2,\dots,W^N$\}, of size $B$ bits each. There are $K$ users in the system, and the number of files in the library exceeds the number of users, i.e., $N\geq K$. The users are assisted by $\Lambda (\leq K)$ helper caches, each with storage capacity of $M$ files. Further, user $k$ has a cache of normalized size $M_k, \  \forall k\in [K]$.

Like the system model introduced in \cite{parrinello2020fundamental}, the $K$ users are partitioned into $\Lambda$ groups, based on the cache they have access to. This is reflected by the \emph{user-to-cache association} $\mathcal{U}=\{\mathcal{U}_1,\mathcal{U}_2,\ldots,\mathcal{U}_\Lambda\}$, where $\forall \lambda \in [\Lambda]=\{1,2,\dots,\Lambda\}$,  $\mathcal{U}_\lambda$ is the ordered set of users accessing cache $\lambda$. The $j^{th}$ user in $\mathcal{U}_\lambda$ is denoted by $\mathcal{U}_\lambda(j)$. The \emph{user-association profile} is given by the vector $\boldsymbol{\mathcal{L}}=(\mathcal{L}_1, \mathcal{L}_2,\ldots,\mathcal{L}_\Lambda)$, with $\mathcal{L}_1\geq\mathcal{L}_2,\ldots,\geq\mathcal{L}_\Lambda$, where $\mathcal{L}_\lambda$ represents the number of users associated to the $\lambda^{th}$ most populated helper cache. The most populated cache has the maximum number of users assigned to it. $\sum_{\lambda=1}^\Lambda\mathcal{L}_\lambda=K$. 
Without loss of generality, we assume that caches are labelled in a non-increasing order of the number of users that access them, i.e., $|\mathcal{U}_\lambda|=\mathcal{L}_\lambda \ \forall \lambda \in [\Lambda]$. Therefore, cache $1$ is accessed by $\mathcal{L}_1$ number of users, cache $2$ is accessed by $\mathcal{L}_2$ number of users, and so on. If not, we can simply relabel the caches to obtain this ordering. 
The system in our formulation operates in four phases: 
\begin{enumerate}[a)]
\item \textit{Helper Cache Placement Phase:} The server, after encoding each file into \emph{shares} populates the helper caches with these shares without knowing the user-to-cache association.
\item \textit{User-to-Cache Assignment Phase:} This occurs after the shares have been placed. Each user is assigned to a single helper cache and $\mathcal{U}$ is communicated to the server.
\item \textit{User Cache Placement Phase:} Once the user-to-cache association is conveyed, the server privately places \emph{unique keys} in the user caches.
\item \textit{Delivery Phase:} Each user requests for a single file from the library. In response, the server broadcasts coded multicast messages over the shared link so as to satisfy the users' demands, and ensuring that the users gain no information about the content of the files they did not request.
\end{enumerate}
To summarize, the system model considered in this paper differs from the one in \cite{parrinello2020fundamental} in the following aspects: 
\begin{itemize}
    \item User $k$ has a dedicated cache of size $M_k\geq 1, \ \forall k\in [K]$ to store keys.
    \item There is an additional phase for placing these keys in the user caches.
\end{itemize}

The content in the helper and user cache, comprising the shares and keys, that user $k$ has access to is denoted by $\mathcal{Z}_k$, $\forall \ k \in [K]$. In the delivery phase, each user requests for a unique file. Users convey their requests to the server in the form of a demand vector $\mathbf{d}=(d_1,d_2,\dots,d_K)$, which represents that user $k$ has requested $W^{d_k}$. We assume $N\geq K$ and the \emph{worst case demand vector}, i.e., $d_i\neq d_j, \ \forall \ \ i,j\in [K], i\neq j$. Let, for a profile $\boldsymbol{\mathcal{L}}$, $X_{\mathbf{d}}(\boldsymbol{\mathcal{L}})$ denote the transmitted message vector corresponding to the demand $\mathbf{d}$. 
Given this setting, the \emph{perfect secrecy} constraint has to be met by all the users. By the definition in \cite{ravindrakumar2018private}, for perfect secrecy, \emph{information leakage} must be $0$, i.e.,
\begin{equation}\label{sec_per}
  I(W^{[N]\setminus d_k};\mathcal{Z}_k,X_\mathbf{d}(\boldsymbol{\mathcal{L}}))=0,\ \forall \ \mathbf{d}\in[N]^K,\forall \  k\in[K].  
\end{equation}
where $I(;)$ is the mutual information, and $W^{[N]\setminus d_k}$ is the set of all files except the one requested by user $k$. The maximum load on the shared link of a shared cache system with profile $\boldsymbol{\mathcal{L}}$, under the constraint of $\eqref{sec_per}$ is given by the \emph{effective number of files} that the server broadcasts in the delivery phase to meet the users' demands when they are all distinct. This is quantified by the number of bits transmitted by the server, normalised by file size, for worst case demand vector and is known as the \emph{secretively achievable rate} $R^s(\boldsymbol{\mathcal{L}})$, expressed in units of files.
\section{Proposed Scheme}
\label{sec3}
Before describing the general scheme, we see an example.

\emph{Example 1:}  Consider a network with $N=K=M=3, M_k\geq 1, k\in[3]$, and file size of $B$ bits. $\mathcal{U}_1=\{1,2\}$, $\mathcal{U}_2=\{3\}$; hence $\boldsymbol{\mathcal{L}}=(2,1)$. The server selects $V^1,V^2,V^3$ randomly and uniformly from the Galois field $\mathbb{F}_{2^{B}}$. In the helper-cache placement phase, the server places, $\forall n \in [3]$, $S^n_1=W^n\oplus V^n$ in cache 1 and $S^n_2=V^n$ in cache 2. The server also selects two random keys $T_{\{1,3\}}$ and  $T_{\{2\}}$ uniformly from $\mathbb{F}_{2^{B}}$. In the user cache placement phase, the server privately places $T_{\{1,3\}}$ in user caches 1 and 3, $T_{\{2\}}$ in user cache 2. In the delivery phase, let $\mathbf{d}=(1,2,3)$. The server transmits $S^1_2\oplus S^3_1\oplus T_{\{1,3\}}$ and $S^2_1 \oplus T_{\{2\}}$. User 1 can decode $W^1$ but obtains no information about $W^2$ and $W^3$. This holds for users 2 and 3 as well. The rate is thus $2$ units.

We now describe the general coded caching scheme and derive the secretively achievable rate $R^s(\boldsymbol{\mathcal{L}})$ for memory points $M$ where $t=\frac{\Lambda M}{M+N}\in\{0,1,\dots,\Lambda-1\}$ is the caching parameter. The rate corresponding to non-integer values of $t$ are achievable through memory sharing. The steps involved are as follows:
\subsection{When $t\in \{1,2,\dots,\Lambda-1\}$}
\label{sec3A}
\subsubsection{File encoding}\label{sec3A1}The first step is to encode each file  in the library using a $\big(\binom{\Lambda-1}{t-1},\binom{\Lambda}{t}\big)$ \emph{non- perfect} secret sharing scheme \cite{farras2016onti}. To do this, the file $W^n, \ \forall n \in [N]$ is split into $(\binom{\Lambda}{t}-\binom{\Lambda-1}{t-1})=\binom{\Lambda-1}{t}$ equal-sized sub-files. The size of each sub-file is $F_s=\frac{B}{\binom{\Lambda-1}{t}}$, and the sub-files of $W^n$ is denoted by the column vector $W^n=(W^n_p),\ p\in \big[\binom{\Lambda-1}{t}\big]$. Further, $\binom{\Lambda-1}{t-1}$ \emph{encryption keys} each of length $F_s$ are generated uniformly and independent of the files from $\mathbb{F}_{2^{F_s}}$. They are denoted by the column vector $V^n=(V^n_q), \ q\in \big[\binom{\Lambda-1}{t-1}\big]$. The idea is to concatenate $V^n$, below $W^n$ and multiply the resulting $\binom{\Lambda}{t}$-length vector, $[W^n;V^n]$ with a Cauchy matrix, as presented in \cite{ma2019secure}. The share vector $S^n=(S^n_r),r\in\big[\binom{\Lambda}{t}\big]$ is given by:
\begin{equation}
    S^n=\mathbf{G}\cdot [W^n;V^n] ,\ \ \forall n\in [N].
\end{equation}
where $\mathbf{G}$ is a $\binom{\Lambda}{t}\times\binom{\Lambda}{t}$-Cauchy matrix over $\mathbb{F}_{2^m}$ with $2^m\geq 2\binom{\Lambda}{t}$ as described in \cite{ma2019secure}. A Cauchy matrix has the property that all its sub-matrices have full rank. This ensures that no information about $W^n$ gets revealed by any subset of $\binom{\Lambda-1}{t-1}$ or less elements of $S^n$. Now, consider the collection of sets $\mathfrak{T}:=\{\mathcal{T}\subset [\Lambda],|\mathcal{T}|=t\}$, where $\mathcal{T}$ represents the indices of helper caches in which $S^n$ is placed. These $\binom{\Lambda}{t}$ subsets are arranged in lexicographic order, thus establishing a one-to-one mapping, $\phi:\mathfrak{T}\mapsto \big[\binom{\Lambda}{t+1}\big]$. Assign $S^n_{\mathcal{T}}\gets S^n_i$ where $i=\phi(\mathcal{T})$. Then, the following equations are satisfied:
\begin{enumerate}[(i)]
\item $H(W^n|S^n)=0.$ 
\item $H(W^n|S^n_\mathcal{T})=H(W^n)\implies I(W^n;S^n_\mathcal{T})=0.$
\end{enumerate}
(i) implies that, given all the $\binom{\Lambda}{t}$ shares, a file can be completely recovered. (ii) implies that any collection of $\binom{\Lambda-1}{t-1}$ shares of a given file reveal no information about that file.
\subsubsection{Placement of shares} The helper caches are filled with these shares over a private link between the caches and the server. Similar to the idea of placing a sub-file in a cache, as in \cite{maddah2014fundamental}, \cite{parrinello2020fundamental}, a share is placed in a cache, if the cache index $\lambda\in\mathcal{T}$. Precisely, the following content is cached in cache $\lambda$:
\begin{equation*}
Z_\lambda=\{S^n_\mathcal{T}:\lambda\in\mathcal{T}, \forall\mbox{ }n \in [N]\}.
\end{equation*}
The memory occupied at each helper cache is therefore, $\frac{N\binom{\Lambda-1}{t-1}B}{\binom{\Lambda-1}{t}}=\frac{Nt}{\Lambda-t}B$ bits. Thus, $M=\frac{Nt}{\Lambda-t}$.
 \subsubsection{User to cache assignment} After the caches are filled, the user to cache association is revealed to the server in the \emph{user to cache assignment} phase. This knowledge is used by the server to design the subsequent steps \emph{4)-5)}.
\subsubsection{Placement of keys} \label{sec3A4} At this stage, the server knows $\mathcal{U}$, and hence derives the profile $\boldsymbol{\mathcal{L}}$. It generates $\sum_{r=1}^{\Lambda-t}\mathcal{L}_r\binom{\Lambda-r}{t}$ unique keys, uniformly from $\mathbb{F}_{2^{F_s}}$ and independent of the files and shares. As indicated earlier, each multicast message is XOR'ed with a key. Hence, the group of users getting served by this multicast transmission needs to store this key privately. Note that, to take advantage of the overlap in shares placed at the helper caches, all users connected to different helper caches should be served in a given round of transmission \cite{parrinello2020fundamental}. To achieve this, $\mathcal{U}_1(1), \mathcal{U}_2(1),\ldots \mathcal{U}_\Lambda(1)$ are served in the first round; $\mathcal{U}_1(2), \mathcal{U}_2(2),\ldots, \mathcal{U}_\alpha(2)$ are served in the second round, where $\alpha=\max_{\lambda\in[\Lambda ]}\{\lambda:2 \leq \mathcal{L}_\lambda\}$ and so on. This way, there are $\mathcal{L}_1$ rounds, and the set of users getting served in round $j$ is $\mathcal{R}_j=\bigcup_{\lambda\in[\Lambda]}\{\mathcal{U}_{\lambda}(j):j\leq\mathcal{L}_\lambda\}$. Consider the set $\mathfrak{Q}:=\{\mathcal{Q}\subseteq[\Lambda]:|\mathcal{Q}|=t+1\}$. For each $j\in[\mathcal{L}_1]$ and $\mathcal{Q}\in \mathfrak{Q}$, determine:
\begin{equation}
\label{chiq}
\chi_{\mathcal{Q}}=\bigcup_{\lambda\in\mathcal{Q}}\{\mathcal{U}_\lambda(j):j\leq\mathcal{L}_\lambda\}.
\end{equation}
 Due to the non-uniformity in $\boldsymbol{\mathcal{L}}$, more than one $\mathcal{Q}$ may yield the same $\chi_{\mathcal{Q}}$ (see \emph{Example 2}) in a given round. Let  $l_{\chi_\mathcal{Q}}$ denote the number of times that $\chi_{\mathcal{Q}}$ appears. 
 The server delivers a unique multicast message for every occurrence of $\chi_{\mathcal{Q}}$, each XOR'ed with a unique key. This key is denoted by $T^{l}_{\chi_{\mathcal{Q}}}$, where $l\in [l_{\chi_\mathcal{Q}}]$. For simplicity of notation, we omit the superscript $l$ from $T^{l}_{\chi_{\mathcal{Q}}}$ if $l_{\chi_\mathcal{Q}}=1$. This gives:
\begin{equation*}
\mathcal{Z}_k=\{S^n_\mathcal{T}:\lambda\in\mathcal{T}\,\forall\,n\in[N]\} \bigcup \{T^{l}_{\chi_{\mathcal{Q}}} \ \forall l\in [l_{\chi_\mathcal{Q}}], k\in\chi_\mathcal{Q}\}.
\end{equation*}
To ensure decodability, the memory occupied by the keys in a user cache should be equal to the number of missing shares that the user needs to reconstruct the requested file. This is equal to $\binom{\Lambda}{t}-\binom{\Lambda-1}{t-1}=\binom{\Lambda-1}{t}=\frac{B}{F_s}$ messages, each message being of $F_s$ bits. Hence, the memory of atleast $B$ bits or 1 file is needed at every user cache to guarantee secrecy requirement.
\subsubsection{Transmission of messages}\label{sec3A5}Having received the demand vector $\mathbf{d}$, the server transmits in $\mathcal{L}_1$ rounds. Each round can be viewed as a dedicated caching network \cite{maddah2014fundamental} with $\Lambda$ users and caching parameter $t$. Hence the transmission scheme involves creating  $\binom{\Lambda}{t+1}$ subsets $\mathcal{Q}$ of caches, each corresponding to a multicast message. For each $\mathcal{Q}$, we create the set of users $\chi_\mathcal{Q}$ as defined in \eqref{chiq}. However, in round $j$, only  users in $\mathcal{R}_j$ are involved, hence some of the $\chi_\mathcal{Q}$ are empty. No transmission occurs if $\chi_\mathcal{Q}=\phi$. If $\chi_{\mathcal{Q}}$ is non-empty, then it corresponds to a transmission formed by XOR'ing the shares and the unique key, $T^{l}_{\chi_{\mathcal{Q}}} \forall l\in [l_{\chi_\mathcal{Q}}]$  shared by the users in $\chi_{\mathcal{Q}}$ which is:
\begin{equation}
x^{l}_{\chi_{\mathcal{Q}}}=\bigoplus_{\lambda\in\mathcal{Q}:\mathcal{U}_\lambda(j)\in\chi_\mathcal{Q}}S^{d_{\mathcal{U}_\lambda(j)}}_{\mathcal{Q}\setminus \{\lambda\}}\oplus T^{l}_{\chi_{\mathcal{Q}}}.
\end{equation}
Therefore, combining all rounds, the overall transmitted message vector for profile $\boldsymbol{\mathcal{L}}$ is 
\begin{equation}\label{transmit}
X_\mathbf{d}(\boldsymbol{\mathcal{L}})=\bigcup_{j\in[\mathcal{L}_1],\mathcal{Q}\subseteq[\Lambda]:|\mathcal{Q}|=t+1} \bigg(\bigcup_{l\in [l_{\chi_\mathcal{Q}}]}x^{l}_{\chi_\mathcal{Q}}\bigg).
\end{equation}
\subsection{When $t=0$}
In this case, the helper caches do not prefetch any content, and the user to cache assignment phase is not required. The server randomly generates $K$ unique keys, each having the size of a file. We denote the keys as $T_k, k\in[K]$ and store $T_k$ at user cache $k$. Once the demands are revealed, the requested files are simply transmitted by XOR'ing each file with the key corresponding to the user who demanded it. Thus, $X_{\mathbf{d}}=W^{d_k}\oplus T_k, \forall k \in [K]$. The worst case secretively achievable rate for $t=M=0$, is given by $R^s(\boldsymbol{\mathcal{L}})=K$, for all $\boldsymbol{\mathcal{L}}$. 
\subsection{Calculation of Rate}
From Section \ref{sec3A5}, note that the server omits the transmissions corresponding to empty $\chi_\mathcal{Q}$, which for round $j$ are $\binom{\Lambda-|\mathcal{R}_j|}{t+1}$. Thus, the effective number of transmissions is:
\begin{equation}
    \label{eq1}
    \binom{\Lambda}{t+1}-\binom{\Lambda-|\mathcal{R}_j|}{t+1}, \ \forall j\in[\mathcal{L}_1],
\end{equation}
each of which is the size of a share $F_s$ bits. On summing over all the $\mathcal{L}_1$ rounds, and normalizing by the file size $B$ bits, the secretively achievable rate is:
\begin{equation}
    \label{eq2}
    R^s(\boldsymbol{\mathcal{L}})=\frac{\binom{\Lambda-|\mathcal{R}_j|}{t+1}}{\binom{\Lambda-1}{t}}\stackrel{a}{=}\frac{\sum_{r=1}^{\Lambda-t}\mathcal{L}_r\binom{\Lambda-r}{t}}{\binom{\Lambda-1}{t}}.
\end{equation} where $\stackrel{a}{=}$ is obtained on performing the steps in (91), Appendix-G of \cite{parrinello2020fundamental}.

\emph{Example 2:}
 Consider a network with $N=8, K=8,\Lambda=4$, $M=8$ and $M_k\geq 1, \  \forall k \in [8]$; $t=\frac{\Lambda M}{M+N}=2$; $\mathcal{U}_1=\{1,2,3\}$, $\mathcal{U}_2=\{4,5\}$, $\mathcal{U}_3=\{6,7\}$, $\mathcal{U}_4=\{8\}$. Hence, $\boldsymbol{\mathcal{L}}=(3,2,2,1)$.\\
\textit{Placement:} The files $W^1,\dots,W^8$ are each encoded using a $(3,6)$ non-perfect secret sharing scheme to generate shares $S^n_{1,2},S^n_{1,3},S^n_{1,4},S^n_{2,3},S^n_{2,4},S^n_{3,4}$ $\forall n\in [8]$, with size $F_s={1/3}^{rd}$ of file size. The shares and keys accessible to the users are:
\begin{align*}
    \mathcal{Z}_1  &=\{S^n_{1,2}, S^n_{1,3}, S^n_{1,4}, T_{\{1,4,6\}}, T_{\{1,4,8\}}, T_{\{1,6,8\}}\}.\\
    \mathcal{Z}_2  &=\{S^n_{1,2}, S^n_{1,3}, S^n_{1,4}, T_{\{2,5,7\}}, T_{\{2,5\}}, T_{\{2,7\}}\}.\\
    \mathcal{Z}_3  &=\{S^n_{1,2}, S^n_{1,3}, S^n_{1,4}, T^1_{\{3\}}, T^2_{\{3\}}, T^3_{\{3\}}\}.\\
    \mathcal{Z}_4  &=\{S^n_{1,2}, S^n_{2,3}, S^n_{2,4}, T_{\{1,4,6\}}, T_{\{1,4,8\}}, T_{\{4,6,8\}}\}.\\
    \mathcal{Z}_5  &=\{S^n_{1,2}, S^n_{2,3}, S^n_{2,4}, T_{\{2,5,7\}}, T_{\{2,5\}}, T_{\{5,7\}}\}.\\
    \mathcal{Z}_6  &=\{S^n_{1,3}, S^n_{2,3}, S^n_{3,4}, T_{\{1,4,6\}}, T_{\{1,4,6\}}, T_{\{1,6,8\}}\}.\\
    \mathcal{Z}_7  &=\{S^n_{1,3}, S^n_{2,3}, S^n_{3,4}, T_{\{2,5,7\}}, T_{\{2,7\}}, T_{\{5,7\}}\}.\\
    \mathcal{Z}_8  &=\{S^n_{1,4}, S^n_{2,4}, S^n_{3,4}, T_{\{1,4,8\}}, T_{\{1,4,8\}}, T_{\{1,6,8\}}\}.
\end{align*}
\textit{Delivery:} Let the demand vector $\mathbf{d}=(1,2,3,4,5,6,7,8)$. 
Since, $\mathcal{L}_1=3$, there are \emph{three} rounds of transmission. The multicast messages are listed in Table \ref{table_1} wherein $j$ refers to the round of transmission.
\begin{table}[htbp]
    \caption{Server transmissions}
    \label{table_1}
    \centering
    \includegraphics[width=0.45\textwidth]{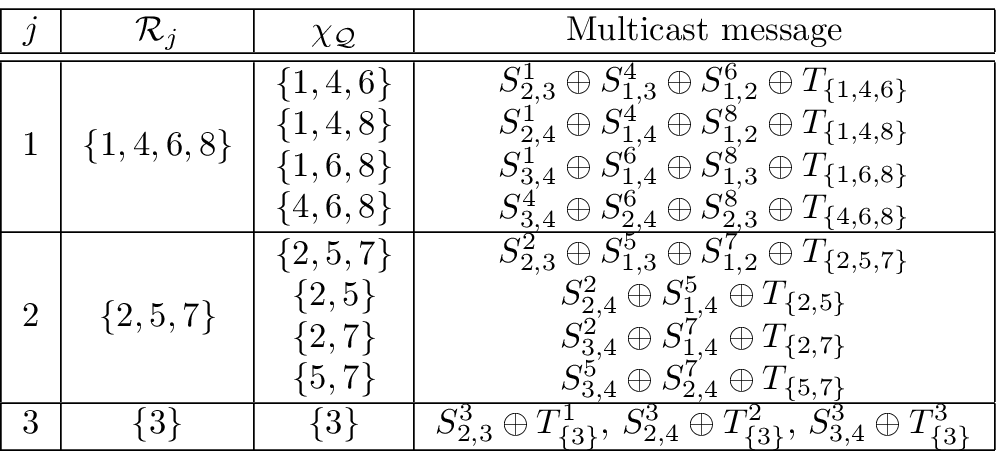}
\end{table}
 Using these $11$ transmissions, the demands of all the users are satisfied. Since each transmission comprises $B/3$ bits, the secretively achievable rate is $R^s(\boldsymbol{\mathcal{L}})=R^s((3,2,2,1))=11/3$ units.
\subsection{Proof of Correctness} Each transmitted message serves the users in the corresponding set $\chi_\mathcal{Q}$ who have the key  $T^l_{\chi_\mathcal{Q}}, l\in [l_{\chi_\mathcal{Q}}]$. As a result, they can recover a linear combination of shares after XOR'ing the received message with the key. In this linear combination, except the missing share of the requested file, all other shares are available to the user via the helper cache. On receiving $\binom{\Lambda-1}{t-1}$ such distinct transmissions, the user can obtain  the missing shares. The reconstruction of the requested file (secret) is possible because all the $\binom{\Lambda}{t}$ shares are now available to the user and the Cauchy matrix $\mathbf{G}$ is invertible.
\subsection{Proof of Secrecy} By the structure of the $\big(\binom{\Lambda-1}{t-1},\binom{\Lambda}{t}\big)$ secret-sharing encoding, the users obtain no information about the files from the helper caches. Furthermore, the unique keys stored are also independent of the files. Thus, there is no information leakage from the caches. For each transmission, the users who do not have the key, cannot decipher the message. Therefore, users connected to a given helper cache, cannot eavesdrop on the messages meant for their neighbour connected to the same helper cache. Thus, \eqref{sec_per} is satisfied.
 \section{Main Results}
\label{sec4}
\begin{thm}
In the $K$- user shared link broadcast channel with $\Lambda$ helper caches and $K$ user caches, of normalized size $M$ and $M_k= 1 \ \forall k\in [K]$ respectively, with $t=\frac{\Lambda M}{M+N}\in\{0,1,\ldots,\Lambda-1\}$ the following rate $R^s(\boldsymbol{\mathcal{L}})$ is secretively achievable within a profile $\boldsymbol{\mathcal{L}}$:
\begin{equation}\label{shd_scc}
R^s(\boldsymbol{\mathcal{L}})=\frac{\sum_{r=1}^{\Lambda-t}\mathcal{L}_r\binom{\Lambda-r}{t}}{\binom{\Lambda-1}{t}}.
\end{equation}
For general $M\in[0,{N(\Lambda-1)}]$, the lower convex envelope of these points is secretively achievable, through memory sharing.
\end{thm}
\begin{IEEEproof}
The proof follows from the proposed scheme in Section \ref{sec3} and the rate calculation in \eqref{eq1} and \eqref{eq2}.
\end{IEEEproof}
\begin{rem}
User cache size $M_k$ of $1$ unit $\forall k\in [K]$ is sufficient for the feasibility of our scheme. If $M_k>1$, the remaining memory stays unoccupied, and the scheme continues to work. If $M_k<1$, the keys required by user $k$ cannot be stored and the multicast messages useful to the user cannot be decoded. To ensure decodability without storing the keys, the messages need to be transmitted without XOR'ing them with keys, which violates \eqref{sec_per}. As long as all user caches have $M_k\geq 1$, our scheme continues to work.
\end{rem}

\begin{figure}[htbp]
    \centering
    \includegraphics[scale=0.55]{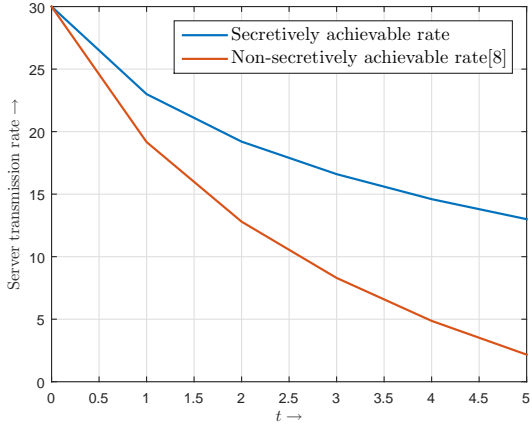}
    \caption{Transmission rates with and without secrecy for shared cache setup with $N=K=30$ and $\Lambda=6$ for profile $\boldsymbol{\mathcal{L}}=(13,8,4,2,2,1)$}
    \label{fig_2}
\end{figure}

In the context of secretive (non-secretive) coded caching, the caching parameter, $t$ indicates the number of helper caches in which a given share (sub-file) is placed. Fig. \ref{fig_2} depicts the rate vs $t$ plots for shared cache setting with and without secrecy imposed. The secretively achievable rate is higher than the non-secretively achievable rate $\forall t$ except when $t=0$. This is because, the former involves \emph{shares}, instead of sub-files in the placement and delivery phases. The shares consume larger number of bits compared to sub-files due to the randomness introduced in the form of encryption keys.
\begin{rem}
For uniform user association profile, where $\Lambda\vert K$, denoted by $\boldsymbol{\mathcal{L}_{\text{unif}}}=(\frac{K}{\Lambda},\frac{K}{\Lambda},\ldots,\frac{K}{\Lambda})$, the secretively achievable rate is obtained by substituting  $\mathcal{L}_r$ in \eqref{shd_scc} as:
\begin{equation}
\label{unif_rate}
  R^s(\boldsymbol{\mathcal{L}}_{\text{unif}})\ =\frac{\frac{K}{\Lambda}\sum_{r=1}^{\Lambda-t}\binom{\Lambda-r}{t}}{\binom{\Lambda-1}{t}}\stackrel{*}{=}\frac{\frac{K}{\Lambda}\binom{\Lambda}{t+1}}{\binom{\Lambda-1}{t}}=\frac{K}{t+1}.
\end{equation}
where $\stackrel{*}{=}$ follows from the Hockey-stick identity.
Further, when $\Lambda=K$, and $\boldsymbol{\mathcal{L}_{\text{MN}}}=(1,1,\dots,1)$, the secretively achievable rate coincides with that in \cite{ravindrakumar2018private}.
\end{rem}
\begin{IEEEproof}
The proof of the second part of the remark is as follows. In the general achievable scheme presented in \cite{ravindrakumar2018private}, the secretively achievable rate $R^s(M)$ is given by:
\begin{equation}
\label{eq_RK}
R^s(M)=\frac{K(N+M-1)}{N+(M-1)(K+1)}.
\end{equation}
In a shared cache system, when the profile is $\boldsymbol{\mathcal{L}_{\text{MN}}}$, each user has a distinct helper cache of size $M$ units and the size of the user cache occupied by keys as $1$ unit. This is equivalent to a dedicated cache network where each user has access to a local cache of $M+1$ units. Hence, $R^s(\boldsymbol{\mathcal{L}_{\text{MN}}})=\frac{K(N+M)}{N+M(K+1)}$ obtained by substituting $t=\frac{KM}{M+N}$ in \eqref{unif_rate} matches the secretively achievable rate $R^s(M+1)$ obtained by setting $M\leftarrow M+1$ in \eqref{eq_RK}.
\end{IEEEproof}
\begin{rem}
When all the users are connected to the same cache, meaning $\boldsymbol{\mathcal{L}}=(K,0,\dots,0)$, then the secretively achievable rate is always $K$, irrespective of the helper cache size. Hence, the purpose of caching is defeated.
\end{rem}
\begin{rem}
The secretively achievable scheme exhibits minimum transmission rate for uniform profile, i.e., when $\boldsymbol{\mathcal{L}=\mathcal{L}_{\text{unif}}}$. As illustrated in Fig. \ref{fig_3}, the rate with $\boldsymbol{\mathcal{L}}=(5,5,5,5,5,5)$ is lower than any non-uniform profile. 
\end{rem}
 \begin{IEEEproof}
 The proof follows directly from Corollary 1 of \cite{parrinello2020fundamental}.
 \begin{figure}[tbp]
    \centering
    \includegraphics[scale=0.55]{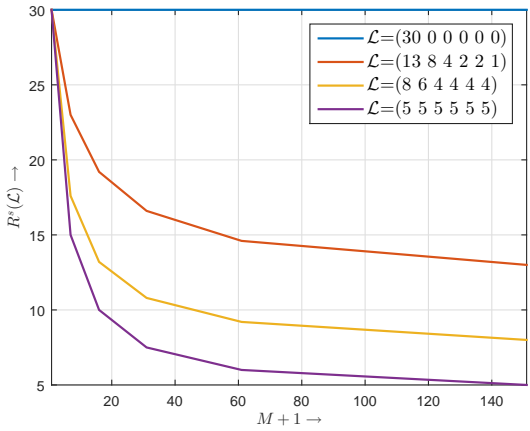}
    \caption{Secretively achievable rate $R^s(\boldsymbol{\mathcal{L}})$ vs memory required $M+1$ (helper cache + user cache occupied with keys) for different user association profiles with $N=K=30$ and $\Lambda=6.$}
    \label{fig_3}
\end{figure}
\end{IEEEproof}
Intuitively, the total number of missing shares is the same irrespective of $\boldsymbol{\mathcal{L}}$. But, the number of users served in a single transmission with $\boldsymbol{\mathcal{L}}_{\text{unif}}$ is always $t+1$, where $t$ is the number of helper caches in which a given share is placed. This way, the uniform profile exploits full multicasting gain, thus requiring minimum number of transmissions. 
\section{Conclusion}
\label{sec5}
In this letter, we proposed a secretively achievable coded caching scheme with shared caches, and showed that when our scheme is applied to the dedicated cache network, the general achievable scheme presented in \cite{ravindrakumar2018private} is recovered. In our solution, we indicated that the users associated to the same cache act as potential eavesdroppers, thus violating the secrecy. Therefore, we secure each message in the delivery phase with a unique key. The secretively achievable rate varies with the user association profile $\boldsymbol{\mathcal{L}}$, and is minimum in the case of uniform profile $\boldsymbol{\mathcal{L}_{\text{unif}}}$. However, testing the optimality of our scheme remains open. Finding an information theoretic lower bound on the optimal rate is an interesting direction to pursue.

\ifCLASSOPTIONcaptionsoff
\newpage
\fi

\end{document}